\documentclass[runningheads]{llncs}
\usepackage[T1]{fontenc}
\usepackage{graphicx}
\usepackage{booktabs}
\usepackage{xcolor}
\usepackage{hyperref}
\usepackage{amssymb}
\begin{document}
\title{LessonBench-V1: A Benchmark Dataset for Evaluating AI Lesson Generation Agents}
\authorrunning{Rammuni Silva et al.}
\titlerunning{LessonBench-V1}  % abbreviated title (for running head)
%                                     \toctithele is used
%
\author{Ravidu Suien Rammuni Silva\inst{1} \and Ahmad Lotfi\inst{1} \and Isibor Kennedy Ihianle\inst{1} \and Golnaz Shahtahmassebi\inst{2} \and Jordan J. Bird\inst{1}}
% \tocauthor{Ravidu Suien Rammuni Silva, Ahmad Lotfi, Isibor Kennedy Ihianle, Golnaz Shahtahmassebi, Jordan J. Bird}
%
\institute{Department of Computer Science, Nottingham Trent University, Nottingham, UK \\Corresponding author: \email{ravidu.rammunisilva2024@my.ntu.ac.uk}
\and
Department of Physics and Mathematics, Nottingham Trent University, Nottingham, UK
}

\maketitle              % typeset the header of the contribution
\begin{abstract}
Large Language Model (LLM) based AI educational content generation systems are increasingly being developed, yet no standardised benchmark exists to systematically evaluate them. This study introduces \textbf{LessonBench-V1}, a benchmark dataset comprising 647 human-written lessons paired with LLM-based reverse-engineered lesson plans across 240 STEM topics spanning mathematics, physics, chemistry, and computer science. The lessons are drawn from 97 trusted open sources, including LibreTexts, Brilliant.org and GeeksForGeeks. Each lesson plan is human-reviewed and produced through a pedagogically grounded methodology that synthesises Bloom's Taxonomy, Gagn\'e's Events, Merrill's First Principles, and the 5E Instructional Model. The lesson plans capture 3,620 learning objectives with pedagogical metadata, enabling systematic, reproducible evaluation of lesson-generation AI agents and supporting further research. The study further proposes a three-dimensional evaluation pipeline for use with the dataset.

\keywords{Educational AI  \and Benchmark Dataset \and Lesson Generation}
\end{abstract}
\section{Introduction}
Large Language Models (LLMs) are increasingly used to generate educational content, ranging from tutoring support to full-length explanatory lessons \cite{xu_educhat_2025,macina_mathdial_2023,silva_generative_2025}. A particularly emerging task is \emph{lesson generation}: given a structured, sufficiently detailed lesson plan specifying topics, learning objectives, and preferred pedagogical delivery, an automated AI system produces a complete lesson. However, evaluating the quality of such AI-generated lessons against human expert standards remains an open challenge \cite{silva_generative_2025}, and there exists no standardised benchmark pairs structured lesson plans with corresponding human-written lessons for systematic evaluation.

Existing educational datasets such as MathDial \cite{macina_mathdial_2023}, TutorEval \cite{chevalier_language_2024}, and EduChat \cite{xu_educhat_2025} address similar issues. TheoremExplainBench \cite{ku_theoremexplainagent_2025} provides 240 STEM topic descriptions with difficulty ratings, but reference lessons are not provided. Open educational resources such as LibreTexts \cite{libretexts_libretexts_2026} offer textbook chapters and other high-quality articles, but they lack structured paired lesson plans. These datasets do not capture the full plan-to-lesson pipeline that an AI generation agent would follow.

Addressing these issues, this short paper presents the following two key contributions: 1) \textbf{LessonBench-V1}: The first open-source\footnote{LessonBench-V1 Dataset: \url{https://github.com/SuienS/lesson-bench-v1}} benchmark dataset pairing structured lesson plans with expert-written lessons across 240 STEM topics and four subjects, comprising 647 paired samples from 97 unique educational sources. The dataset is constructed using a reproducible, multi-stage data curation pipeline in collecting, cleaning, structuring, and reviewing educational content from web sources. 2) A pedagogical research-grounded approach to reverse-engineering lesson plans, incorporating Bloom's Taxonomy \cite{anderson_taxonomy_2001}, Gagn\'e's Events \cite{gagne_conditions_1985}, Merrill's First Principles \cite{merrill_first_2002}, and the 5E Model \cite{bybee_bscs_2015} to extract detailed, structured instructional blueprints from existing lessons.

\section{Related Work}
\paragraph{Educational AI Benchmarks.} 
Several benchmarks evaluate AI in educational settings, but none cover the full plan-to-lesson pipeline, as shown in Table \ref{tbl:comparison}. MathDial \cite{macina_mathdial_2023} provides 3,000 teacher-student dialogues for tutoring evaluation, TutorEval \cite{chevalier_language_2024} evaluates LLMs for science tutoring, and EduChat \cite{xu_educhat_2025} focuses on educational chatbot design. TheoremExplainBench \cite{ku_theoremexplainagent_2025} provides 240 STEM topics that form the foundation of this work, but contains no reference lessons.

\paragraph{Instructional Design Frameworks.}
The lesson plans in LessonBench-V1 were constructed by incorporating four established pedagogical frameworks. Bloom's revised Taxonomy \cite{anderson_taxonomy_2001} provides six cognitive levels for tagging learning objectives. Gagn\'e's Nine Events of Instruction \cite{gagne_conditions_1985} informs the instructional flow, together with Merrill's First Principles \cite{merrill_first_2002} and the BSCS 5E Model \cite{bybee_bscs_2015}. LessonBench lesson plans unify these into six instructional roles: Motivation, Activation, Demonstration, Application, Integration and Assessment, which holistically cover the instructional flow.

\paragraph{LLMs for Educational Content.}
Recent work has explored LLMs for the generation of educational materials \cite{silva_generative_2025}. However, most evaluations rely on ad hoc human judgement or LLM-as-judge approaches \cite{zheng_judging_2023} rather than on reproducible, reference-based benchmarking. LessonBench-V1 aims to provide this missing reference dataset to enable systematic and reproducible evaluation.

\begin{table}[t]
\centering
\caption{Comparison with existing educational AI datasets.}
\label{tbl:comparison}
\begin{scriptsize}
\begin{tabular}{@{}lccccc@{}}
\toprule
\textbf{Dataset} & \textbf{Modality} & \textbf{Paired Plan} & \textbf{Subjects} & \textbf{Full Lessons} & \textbf{Scale} \\
\midrule
TheoremExplainBench \cite{ku_theoremexplainagent_2025} & Topics & \texttimes & 4 STEM & \texttimes & 240 topics \\
MathDial \cite{macina_mathdial_2023}            & Dialogues & \texttimes & Math & \texttimes & 3k dialogues \\
TutorEval \cite{chevalier_language_2024}           & Q\&A  & \texttimes & Science & \texttimes & 400 samples \\
LibreTexts \cite{libretexts_libretexts_2026}            & Textbooks & \texttimes & Multiple & Partial & Varies \\
\midrule
\textbf{LessonBench-V1} & \textbf{Full lessons} & \checkmark & \textbf{4 STEM} & \checkmark & \textbf{647 pairs} \\
\bottomrule
\end{tabular}
\end{scriptsize}
\end{table}

\section{Dataset Construction}

\subsection{Source Data and Topic Selection}
Educational topics chosen for this study were drawn from TheoremExplainBench \cite{ku_theoremexplainagent_2025}, which includes 240 topics across Mathematics, Physics, Chemistry, and Computer Science, categorised into three difficulty levels: Easy, Medium, Hard. For each topic, LessonBench-V1 contains 1-5 human-written lessons from various educational sources, totalling 647 lessons. Each lesson is accompanied by a corresponding lesson plan. Source selection prioritised reliable, openly accessible, and pedagogically structured content. The 97 source websites span open textbook platforms such as LibreTexts (47\%), which include resources from a variety of high-quality sources, as well as educational portals such as GeeksForGeeks (9\%), Byjus (8\%), and Brilliant.org (5\%).

\subsection{Multi-Stage Curation Pipeline}

\begin{figure}[t]
    \centering
    \includegraphics[width=1\linewidth]{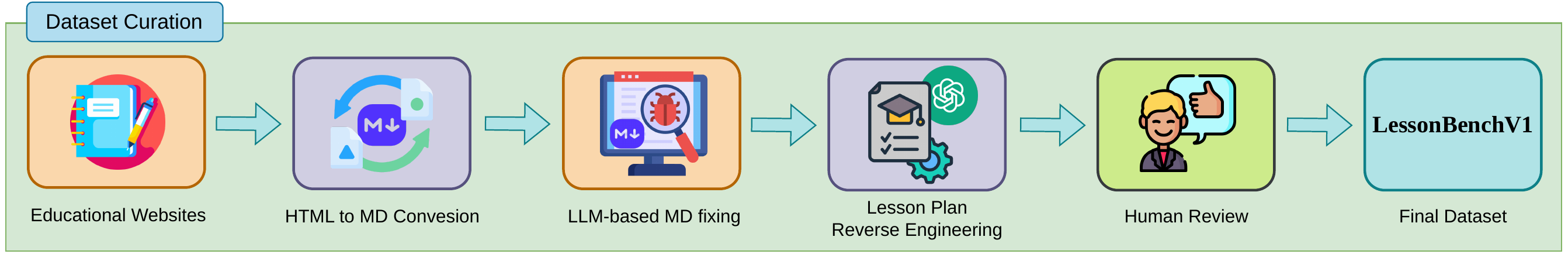}
    \caption{LessonBench construction pipeline}
    \label{fig:lb-construction}
\end{figure}

\begin{enumerate}
    \item \textbf{Web Data Collection:} Relevant sources, including related images, are collected and downloaded.
    \item \textbf{HTML-to-Markdown Conversion:} The downloaded HTML content is converted to Markdown using the \texttt{html-to-markdown} Python library.
    \item \textbf{LLM-based Markdown Cleanup:} The resulting Markdown often contains unwanted web-scraping artefacts, such as navigational text, broken LaTeX equations, and unrelated web-page content. A carefully designed prompt \footnote{MD Fixing system prompt: \url{https://github.com/SuienS/lesson-bench-v1/blob/main/prompts/MD_Cleanup_Prompt.md}} is used with Qwen 3.6 35B A3B to fix formatting issues across ten identified categories while preserving the relevant educational content verbatim, with no rewriting, summarising, or additional content.
    \item \textbf{Lesson Plan Reverse Engineering:} A pedagogically research-backed prompt \footnote{Lesson Plan Reverse Engineering system prompt: \url{https://github.com/SuienS/lesson-bench-v1/blob/main/prompts/LessonPlan_Creation_prompt.md}} was used with GPT-4.1, accessed via the GitHub Copilot API, to reverse engineer a structured lesson from each cleaned lesson. The prompt is designed to holistically `reverse engineer' the instructional blueprint without summarising content. The resulting plan captures six sections: Metadata, Learning Objectives, Key Concepts and Essential Questions, Instructional Flow, Practice and Application Strategy, and Pedagogical Strategy. Each lesson plan\footnote{Sample Lesson Plan: \url{https://github.com/SuienS/lesson-bench-v1/blob/main/LessonBench/data/topic_000/lesson_000_0/lesson_plan/lesson_plan.md}} follows a structured template grounded in four pedagogical frameworks. Learning objectives use Bloom's revised taxonomy verbs \cite{anderson_taxonomy_2001} and are tagged to one of the six cognitive levels. Instructional flow phases are annotated with roles derived from cross-mapping Gagn\'e's Events \cite{gagne_conditions_1985}, Merrill's Principles \cite{merrill_first_2002}, and the 5E model \cite{bybee_bscs_2015} into six refined roles: \emph{Motivation}, \emph{Activation}, \emph{Demonstration}, \emph{Application}, \emph{Integration}, and \emph{Assessment}.
    \item \textbf{Human Review:} The cleaned lesson and the created lesson plan are then reviewed to identify any further formatting or structural issues, which are then fixed manually.
\end{enumerate}

\section{Dataset Analysis}
% \subsection{Summary Statistics}
Table \ref{tbl:dataset-stats} summarises key statistics for LessonBench-V1. The dataset comprises 647 pairs of lessons and lesson plans across 240 topics. Human-written lessons average 1,744 words, whereas lesson plans are substantially shorter, averaging 938 words.

\begin{table}[t]
\centering
\caption{LessonBench dataset summary statistics.}\label{tbl:dataset-stats}
\setlength{\tabcolsep}{5pt}
\begin{tabular}{@{}lr@{}}
\toprule
\textbf{Statistic} & \textbf{Value} \\
\midrule
Total topics & 240 \\
Total lesson and lesson plan pairs & 647 \\
Lessons per topic (mean / median) & 2.70 / 3 \\
Subjects & 4 (Math, Physics, CS, Chemistry) \\
Difficulty levels & 3 (Easy, Medium, Hard) \\
Unique source websites & 97 \\
\midrule
Avg.\ lesson word count & $1{,}744 \pm 1{,}758$ \\
Avg.\ lesson plan word count & $938 \pm 282$ \\
Avg.\ learning objectives per plan & $5.6 \pm 1.1$ \\
Avg.\ instructional flow phases & $7.8 \pm 2.8$ \\
\bottomrule
\end{tabular}
\end{table}

\begin{figure}[t]
    \centering
    \includegraphics[width=1\linewidth]{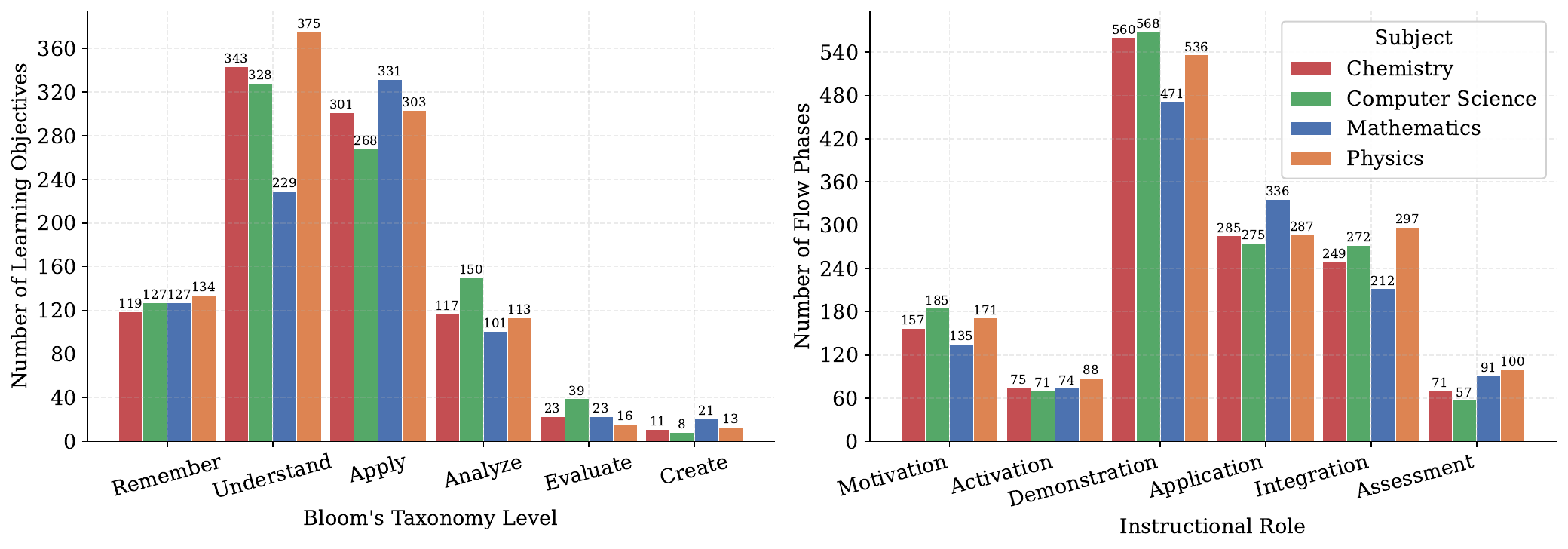}
    \caption{The distribution of Bloom's Taxonomy level (left) and Instructional Roles (right) across subjects}
    \label{fig:bloom-ins-distribution}
\end{figure}

\paragraph{Bloom's Taxonomy Distribution.}
Across all 647 lessons, 3,620 learning objectives were extracted, each assigned a Bloom's taxonomy level, as shown in Figure \ref{fig:bloom-ins-distribution}. 
% Figure \ref{fig:bloom-distribution-grouped} shows the distribution of Bloom's taxonomy levels grouped by topic difficulty.

\paragraph{Instructional Role Distribution.}
The distribution of the 5,623 parsed instructional role annotations is shown in Figure \ref{fig:bloom-ins-distribution}. Subject-specific patterns are evident in the pedagogical metadata. Mathematics lessons have the highest proportion of Apply-level objectives, reflecting an emphasis on procedural problem-solving. Physics lessons show the highest proportion of `Understand' objectives, consistent with conceptual explanation being central to physics pedagogy. Computer Science lessons have the most balanced Bloom's distribution, with notable Analyse-level content focused on algorithm analysis.

% \begin{figure}[t]
%     \centering
%     \includegraphics[width=0.7\linewidth]{role_distribution}
%     \caption{Distribution of the instructional roles grouped by the subject category.}
%     \label{fig:ins-role-dist}
% \end{figure}

\section{Benchmarking Framework}
LessonBench-V1 supports the systematic evaluation of AI lesson-generation systems that take a lesson plan and produce a complete lesson from it. Given a lesson plan $P_i$ from the dataset, an AI system generates a lesson $\hat{L}_i$, which is then compared with the corresponding human-written reference $L_i$. This mirrors the real-world use case in which a curriculum designer writes a plan and an AI agent produces the lesson.

LessonBench can be used across three complementary evaluation dimensions: 1) \emph{Semantic similarity} via BERTScore \cite{zhang2019bertscore}, measuring contextual overlap; 2) \emph{Structural similarity} via ROUGE-L \cite{lin2004rouge}, capturing the longest common subsequence overlap; and 3) \emph{Pedagogical alignment} by reverse-engineering a lesson plan from the generated lesson and comparing Bloom's level coverage and role distributions with the original plan. Furthermore, the datasets' metadata, derived from TheoremExplainBench, enables systematic analysis of these dimensions across aspects such as subject and difficulty.

\section{Conclusion and Future Work}
This short paper presents LessonBench-V1, a benchmark dataset that pairs structured lesson plans with expert-written lessons across 240 topics spanning four STEM subjects, resulting in 647 paired samples. The dataset is collected and constructed using a reproducible five-stage pipeline. Our initial analysis shows meaningful patterns across subjects and difficulty levels in the collected lessons, affirming its validity.

\paragraph{Limitations and Future Work.}
The lesson plans for the dataset were generated using GPT 4.1. Although the generated plans were reviewed prior to the final compilation of the dataset, we plan to conduct systematic experiments to statistically verify their completeness and accuracy against an expert-written lesson plan baseline. The dataset is currently drawn only from English-language sources, and future versions will include lessons from multiple languages. Furthermore, this dataset is an initial step in a larger research effort to develop a full end-to-end lesson-generation AI agent, which is planned to be evaluated on it.

% \begin{credits}
% % \subsubsection{\ackname} A bold run-in heading in small font size at the end of the paper is
% % used for general acknowledgments, for example: This study was funded
% % by X (grant number Y).

% \subsubsection{\discintname}
% The authors have no competing interests to declare that are
% relevant to the content of this article.
% \end{credits}
\bibliographystyle{splncs04}
\bibliography{bibliography}

@book{anderson_taxonomy_2001,
	title = {A taxonomy for learning, teaching, and assessing: {A} revision of {Bloom}'s taxonomy of educational objectives},
	publisher = {Longman},
	author = {Anderson, Lorin W and Krathwohl, David R},
	year = {2001},
}

@book{gagne_conditions_1985,
	edition = {4th},
	title = {The {Conditions} of {Learning} and {Theory} of {Instruction}},
	publisher = {Holt, Rinehart and Winston},
	author = {Gagné, Robert M},
	year = {1985},
}

@article{merrill_first_2002,
	title = {First {Principles} of {Instruction}},
	volume = {50},
	doi = {10.1007/bf02505024},
	journal = {Educational Technology Research and Development},
	author = {Merrill, M. David},
	month = sep,
	year = {2002},
	pages = {43--59},
}

@book{bybee_bscs_2015,
	title = {The {BSCS} {5E} {Instructional} model: {Creating} {Teachable} {Moments}},
	publisher = {National Science Teachers Association},
	author = {Bybee, Rodger W},
	year = {2015},
}

@inproceedings{macina_mathdial_2023,
	title = {Mathdial: {A} dialogue tutoring dataset with rich pedagogical properties grounded in math reasoning problems},
	booktitle = {Findings of the {Association} for {Computational} {Linguistics}: {EMNLP} 2023},
	author = {Macina, Jakub and Daheim, Nico and {others}},
	year = {2023},
	pages = {5602--5621},
}

@misc{silva_generative_2025,
	title = {Generative {Artificial} {Intelligence} and the {Future} of {Education}: {A} {Systematic} {Review} of {Trends}, {Trustworthiness} and {Technological} {Evaluation}},
	copyright = {https://creativecommons.org/licenses/by/4.0/},
	shorttitle = {Generative {Artificial} {Intelligence} and the {Future} of {Education}},
	doi = {10.21203/rs.3.rs-8248369/v1},
	urldate = {2026-06-07},
	publisher = {In Review},
	author = {Silva, Ravidu Suien Rammuni and Lotfi, Ahmad and Ihianle, Isibor Kennedy and Shahtahmassebi, Golnaz and Bird, Jordan J.},
	month = dec,
	year = {2025},
}

@incollection{xu_educhat_2025,
	address = {Singapore},
	title = {{EduChat}: {A} {Large} {Language} {Model}-{Based} {Conversational} {Agent} for {Intelligent} {Education}},
	volume = {2229},
	isbn = {978-981-96-1808-8 978-981-96-1809-5},
	shorttitle = {{EduChat}},
	doi = {10.1007/978-981-96-1809-5_22},
	language = {en},
	urldate = {2026-06-07},
	booktitle = {China {Conference} on {Knowledge} {Graph} and {Semantic} {Computing} and {International} {Joint} {Conference} on {Knowledge} {Graphs}},
	publisher = {Springer Nature Singapore},
	author = {Dan, Yuhao and Lei, Zhikai and Gu, Yiyang and {others}},
	year = {2025},
	pages = {297--308},
}

@article{chevalier_language_2024,
	title = {Language models as science tutors},
	journal = {arXiv preprint arXiv:2402.11111},
	author = {Chevalier, Alexis and Geng, Jiayi and {others}},
	year = {2024},
}

@inproceedings{ku_theoremexplainagent_2025,
	address = {Vienna, Austria},
	title = {{TheoremExplainAgent}: {Towards} {Video}-based {Multimodal} {Explanations} for {LLM} {Theorem} {Understanding}},
	shorttitle = {{TheoremExplainAgent}},
	doi = {10.18653/v1/2025.acl-long.332},
	language = {en},
	urldate = {2026-06-07},
	booktitle = {Proceedings of the 63rd {Annual} {Meeting} of the {Association} for {Computational} {Linguistics} ({Volume} 1: {Long} {Papers})},
	publisher = {Association for Computational Linguistics},
	author = {Ku, Max and Chong, Cheuk Hei and Leung, Jonathan and {others}},
	year = {2025},
	pages = {6663--6684},
}

@misc{libretexts_libretexts_2026,
	title = {{LibreTexts} - {Free} the {Textbook}},
	url = {https://libretexts.org/},
	urldate = {2026-06-07},
	publisher = {Libretexts.org},
	author = {LibreTexts},
	year = {2026},
}

@inproceedings{zheng_judging_2023,
	address = {Red Hook, NY, USA},
	series = {{NIPS} '23},
	title = {Judging {LLM}-as-a-judge with {MT}-bench and {Chatbot} {Arena}},
	booktitle = {Proceedings of the 37th {International} {Conference} on {Neural} {Information} {Processing} {Systems}},
	publisher = {Curran Associates Inc.},
	author = {Zheng, Lianmin and Chiang, Wei-Lin and Sheng, Ying and {others}},
	year = {2023},
}

@article{zhang2019bertscore,
  title={Bertscore: Evaluating text generation with bert},
  author={Zhang, Tianyi and Kishore, Varsha and {others}},
  journal={arXiv preprint arXiv:1904.09675},
  year={2019}
}

@inproceedings{lin2004rouge,
  title={Rouge: A package for automatic evaluation of summaries},
  author={Lin, Chin-Yew},
  booktitle={Text summarization branches out},
  pages={74--81},
  year={2004}
}
\end{document}